# Resonant Inductive Coupling as a Potential Means for Wireless Power Transfer to Printed Spiral Coil

**Mohammad Haerinia**
Dept. of Electrical Engineering
Shahid Beheshti University, Tehran, Iran
m.haerinia@mail.sbu.ac.ir

**Ebrahim S. Afjei**
Dept. of Electrical Engineering
Shahid Beheshti University, Tehran, Iran
e-afjei@sbu.ac.ir

*Abstract*: *This paper proposes an inductive coupled wireless power transfer system that analyses the relationship between induced voltage and distance of resonating inductance in a printed circuit spiral coils. The resonant frequency produced by the circuit model of the proposed receiving and transmitting coils are analysed by simulation and laboratory experiment. The outcome of the two results are compared to verify the validity of the proposed inductive coupling system. Experimental measurements are consistent with simulations over a range of frequencies spanning the resonance.*

## 1. Introduction

Wireless power transfer systems have attracted great attention recently [1]. The power transformation has been electrified for environmental, energy-related and other reasons for many years [2]. Inductive coupling is the most common method of energy transformation [3]. Inductive power transfer is the wireless transformation of electrical power over large distances. Some applications of wireless power transfer include wireless charging and powering implanted electronic devices, medical devices, mobile electronics, implantable medical devices and powered radio way electric vehicles [4-7]. Inductive Power Transfer (IPT) systems have been widely used over the past decade for transmitting tens to hundreds of watts [8]. This technology has proved to be essential [9]. This method has been known as a technique for delivering small amounts of energy to remote devices, however, there are other power transfer methods including infrared and radio frequency (RF) methods which all have their own limitations. The restriction regarding the Infrared technology is requiring direct line of sight between the receiver and the transmitter and RF technology has a low power limitation. A safe and wireless method for transferring energy is inductive coupling [3, 10]. The main drawbacks of this method include the difficult mechanical coupling process between the primary and the secondary side and also the fact that the air gap is less than 60 mm and it is typical high power density [11]. According to the features and drawbacks of other technologies, inductive coupling is preferred for wireless transformation of power [3]. The motivation of this paper is the potential application of magnetic resonant coupling as a way for wireless transformation of power from a source coil to a receiving coil. By optimizing the coil design, the quality factor would be improved [5, 12]. In this work we design the shape of a spiral coil presented in [5] so that it is efficient for transferring high powers. The approximate expressions for the inductance of circular planar inductors have been obtained using accurate expressions for planar spiral inductances which were partially verified by approximately 60 measurements reported in [13]. By analysing the mathematical expression regarding to determining the circuit capacitors and inductors, the resonant coupling in [14] was obtained. The inductive coupled system has been obtained by modifying this procedure for our aim and it is verified with simulation results. The proposed diameters for the coils are 45.2 and 36.4mm and the air gap is considered to vary from 5mm to 15mm. The experimental results were compared with simulations and the design procedure was verified. This work provides the fundamental concept of resonant near-field power transmission and the relation of induced electromotive force and the air gap along the transceiver.

## 2. The concepts of inductive coupling approach

Figure1 shows that inductive coupling-based WPT utilizes magnetic field induction. It is a typical near-field transmission technique.

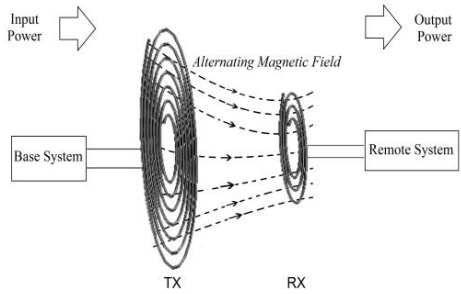

**Fig. 1.** Near-field transmission technique.

Inductive power transfer depends on lots of parameters such as the distance between the coils, frequency, current excitation, the geometry of the coils [15]. The basis of inductive coupling includes the use of the fundamental laws of physics. The calculation of the magnetic field, generated by current distribution throughout coils, can be completed by using Biot-Savart's Law [16].

$$B = \frac{\mu_0}{4\pi} \oint \frac{I d_I \times r}{|r|^3} \quad (1)$$

Where:

$\mu_0$ is the permeability of free space, I is the current in the transmitter coil, $d_I$ is the length of the differential element of the wire, and r is the full displacement vector from the wire element.

By assuming a homogenous flux density distribution in the air gap and neglecting fringing flux, the air gap reluctance can be calculated as:

$$R_g = \frac{l_g}{\mu_0 A_g} \quad (2)$$

Where:

lg, Ag and μ0 are the air gap length, air gap cross section and the permeability of free space, respectively.

The relation (2) is only accurate when the distance between the coils is very small compared to the dimension of the air gap cross-section [17]. The induced voltage over the receiver coil can be computed by using Faraday's Law [16]. The relation can be written as (3):

$$V_{Ind} = -\frac{\partial}{\partial t} \oint B \cdot d_s \quad (3)$$

An inductive power transfer system is comparable to that of a transformer [18]. An alternating current in a transmitting coil generates a varying magnetic field that induces a voltage across the terminals of a receiving coil. Power transmission efficiency is higher when the transmitter coil and the receiver coil are close and aligned. Inductive coupling is a popular and widely utilized technology, with numerous applications, ranging from everything to electric toothbrushes, cell charging, and medical implants [19].

### 3. Design procedure of spiral coil

The proposed coil shown in Fig. 2. Was optimized by evolving the shape of the spiral coils. In this section, the design of the spiral coil is described and coil specifications for the effective permeability are presented.
The variables that parameterize the shape of the coil are the wire thickness (t), inner diameter ($d_{in}$), outer diameter ($d_{out}$), number of turns (N), and the space between turns (s). Transmitting and receiving coil specifications can be seen in Table 1.

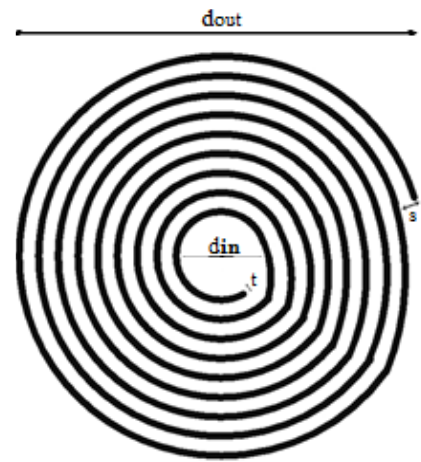

**Fig. 2.** Printed spiral coil.

Several equations have been proposed for approximating L [12]. For designing the mentioned method, we adopted (1) from [13].

$$L = \frac{\mu_0 n^2 d_{avg}}{2}\left[ln\left(\frac{2.46}{\gamma}\right) + 0.20\gamma^2\right] \quad (1)$$

$$\gamma = \frac{d_{out} - d_{in}}{d_{out} + d_{in}} \quad (2)$$

Where n is the number of turns, $d_o$ and $d_i$ are the outer and inner diameters of the coil, respectively.

$d_{avg} = \frac{d_{out}+d_{in}}{2}$ and $\gamma$ is a parameter defined as fill factor [12].

**Table 1.** Specifications of transmitter and receiver.

| Parameters | Transmitter | Receiver |
|---|---|---|
| Inner Diameter (din)(mm) | 10 | 10 |
| Outer Diameter(dout)(mm) | 45.2 | 36.4 |
| Number of turns(N) | 8 | 6 |
| Trace Width(t)(mm) | 0.8 | 0.8 |
| Spacing(s)(mm) | 1.4 | 1.4 |

### 4. Circuit analysis of a resonant coupled system

The operation of an inductive power transfer system can be compared to an air core transformer. The resonant coupling will boost power transfer efficiency, then capacitors, which have been connected to both sides, make a resonant inductive coupling system [18]. Circuit can be classified according to the mode of connection of the capacitor to the coil: Series-Series, Series-Parallel, Parallel-Series and Parallel-Parallel [18, 20], here Series-Parallel mode has been proposed. Simplified schematic of wireless power transfer system using series–parallel transformation network is illustrated in Fig. 3.

In this work, capacitances are chosen such that identical resonant frequency is yielded. As a result, the simple means in (3) are represented so as to achieve resonant coupling along both sides [14, 21]

$$\omega_0 = \frac{1}{\sqrt{L_S C_S}} = \frac{1}{\sqrt{L_P C_P}} \quad (3)$$

For expressing the circuit equations, the following parameters are defined. Component values is shown in Table. 2.

**Table 2.** Component values.

| Elements | Value |
|---|---|
| Cs(nF) | 1.8 |
| Cp(nF) | 3.6 |
| Lt(µH) | 1.589 |
| Lr(µH) | 0.802 |
| Input voltage amplitude(p-p)(V) | 19.6 |
| Resistive load($R_L$)(Ω) | 10 |

### 5. Modelling of transceiver for finite element analysis and simulation results

The finite element method is a useful analysis tool which generates a mesh. The different types of elements produced by the mesh generator are able to be converted to another type. The size of the elements are different from region to region [22]. In this work, two-dimensional Finite Element Method (2-D FEM) COMSOL software is employed to simulate system. The geometry of the transceiver and the corresponding windings as in Fig.4. Is used to simulate the various possible configurations.

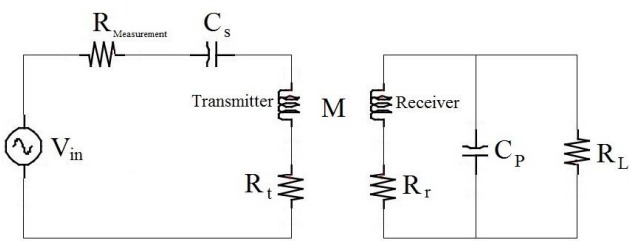

**Fig. 3.** Equivalent circuit of inductive coupling system.

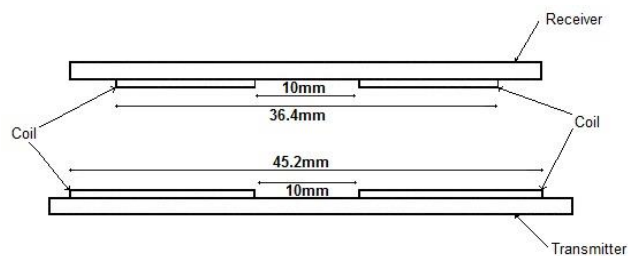

**Fig.4.** Transceiver.

Figures 5-8. Illustrate a set of simulated results of magnetic flux density, magnetic field norm, electric field norm and power flow for proposed model.

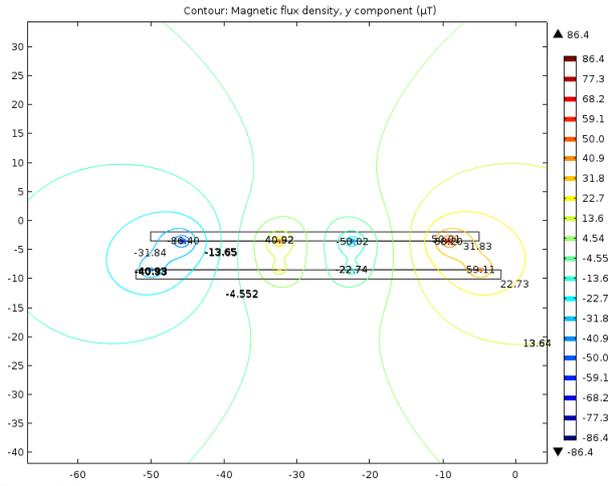

**Fig. 5.** Magnetic flux density norm of proposed transceiver at 3MHz (µT).

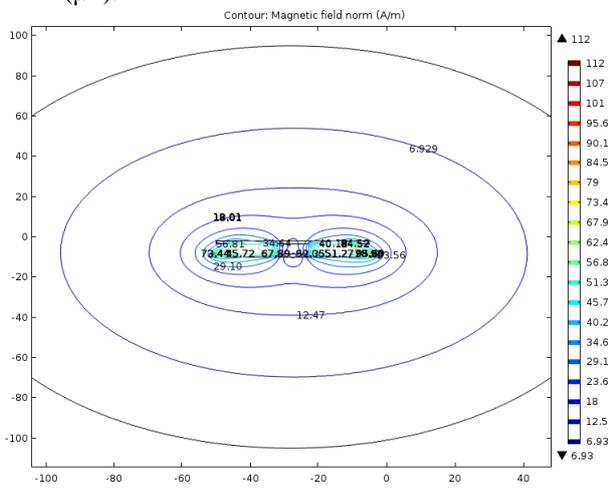

**Fig. 6.** Magnetic field norm of proposed transceiver at 3MHz ($\frac{A}{m}$).

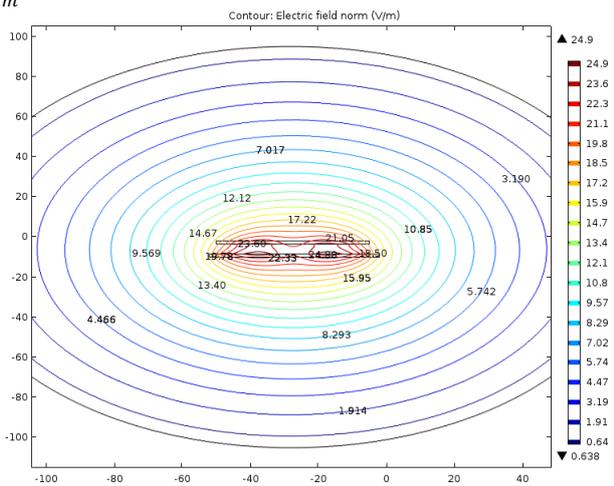

**Fig. 7.** Electric field norm of proposed transceiver at 3MHz ($\frac{V}{m}$).

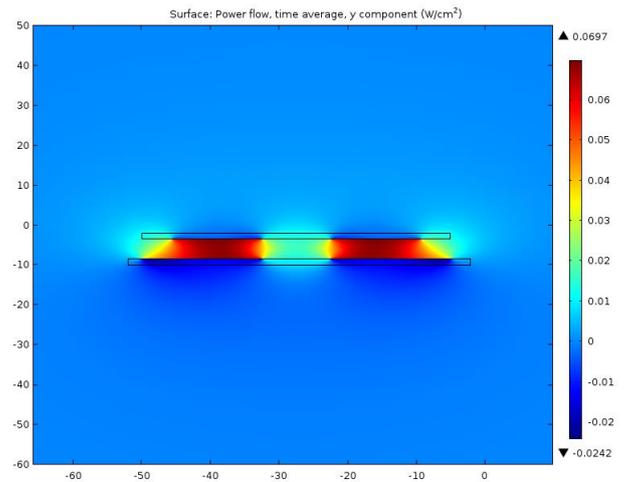

**Fig. 8.** Power density of proposed transceivers, y component at 3MHz ($\frac{W}{cm^2}$).

In this section, a boundary condition has been selected, as Fig.9. For assessment of effect of air gap on power density.

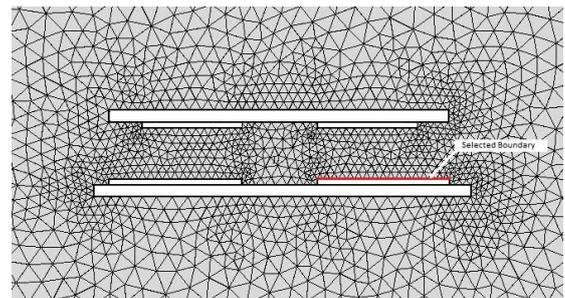

**Fig. 9.** Selected boundary.

Fig.10. Illustrates air gap is changing 5mm to 25mm. as the air gap reluctance increases, power density decreases. Power density will have different magnitude on different parts of the transceiver.

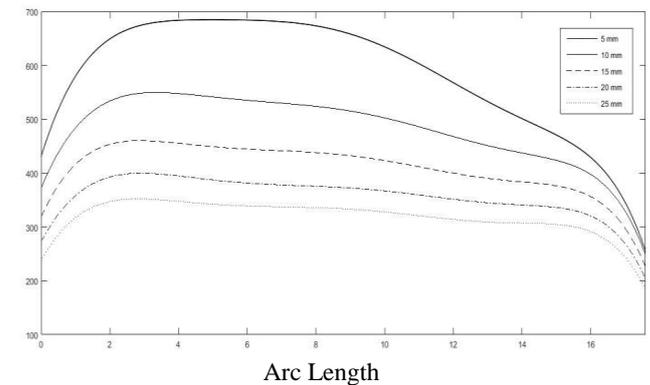

Arc Length

Fig. 10. Power density of proposed transceiver, y component at 3MHz ($\frac{W}{m^2}$).

## 6. Investigation of experimental measurements

The voltage source to the transceiver was provide by function generator with 3 MHz frequency and 19.6 of Vp-p (AC).To record the exact waveform of an electrical signal in this study used oscilloscope connected to software on pc. Fig.11 shows Photograph of the implemented system.

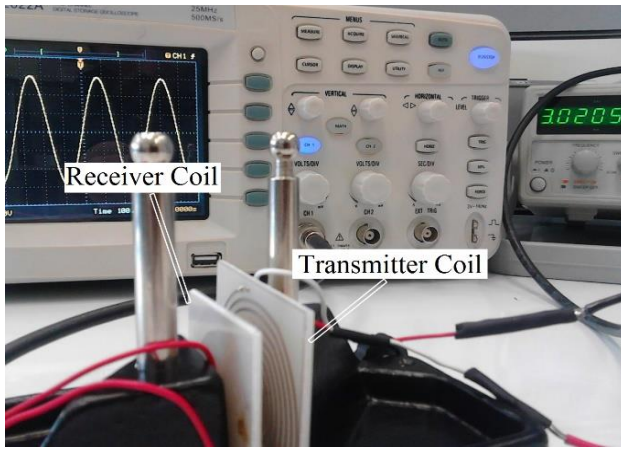

**Fig. 11.** Experimental setup showing the transmitting and receiving coils.

The voltage waveform on the receiver coil were measured at distances from 5mm to 15mm while increasing the distance between the coils. Figure 12 represents Voltage waveform on the transmitter coil. Figures 13 through 15 illustrate a set of measured voltage waveform through the secondary coil changes with distance between the transceiver.

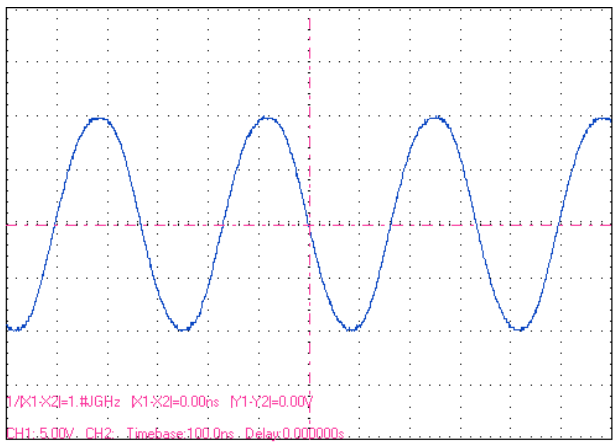

**Fig 12.** Voltage waveform on the transmitter coil.

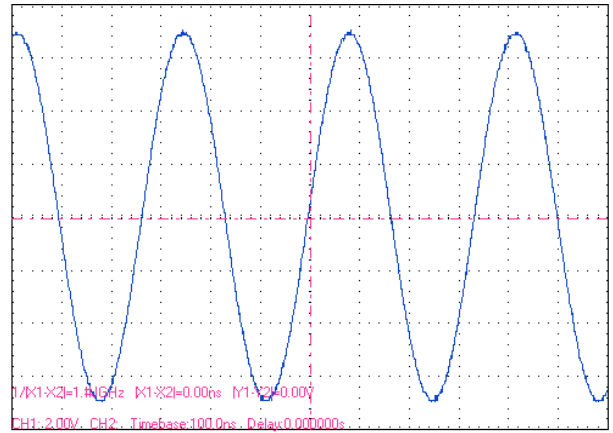

**Fig 13.** Voltage waveform on the receiver coil, at a distance of 5mm.

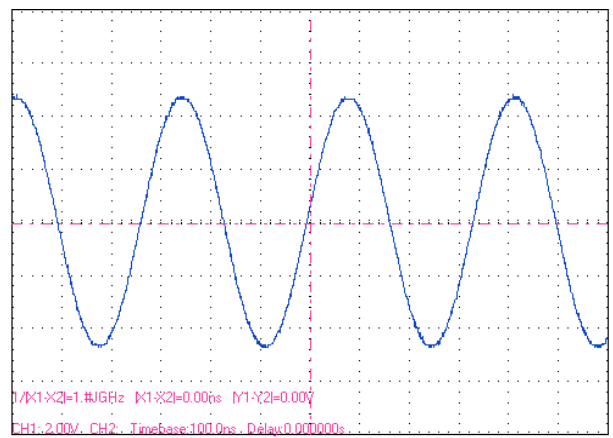

**Fig 14.** Voltage waveform on the receiver coil, at a distance of 10mm.

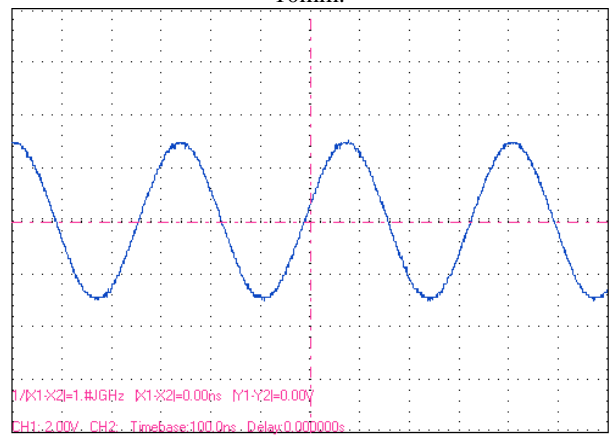

**Fig 15**. Voltage waveform on the receiver coil, at a distance of 15mm.

From the illustrated figures, it's obvious that the wireless power transmission is higher when the distance is nearer. This point was demonstrated by drawing the rms voltage through investigated distances, both from experiments and theory Fig .16.

The simulation model demonstrates the experiments. The voltages were sampled and transferred to MATLAB in order to carry out a curve fitting process. Most of the measured values closely follow the calculated value from MATLAB.

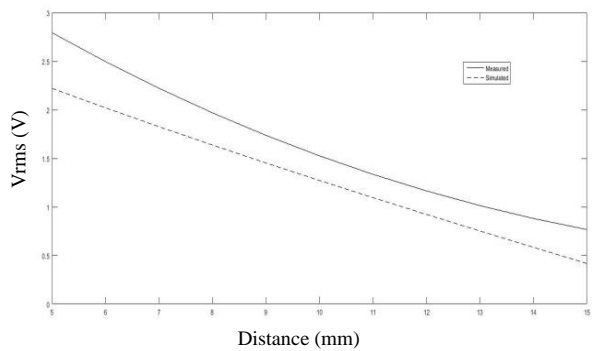

**Fig. 16.** Rms voltage on secondary coil were measured and simulated at distances ranging from 5mm to 15mm.

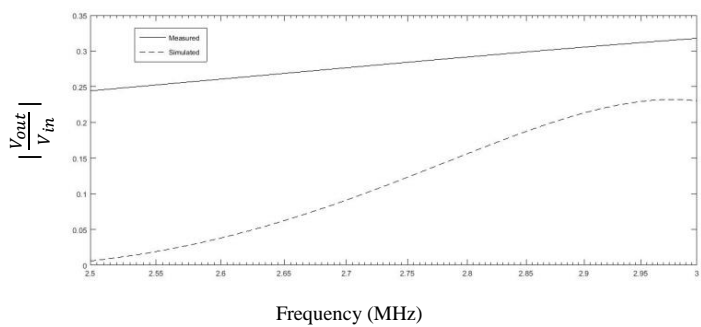

**Fig. 17.** Comparison of transfer function $|\frac{V_{out}}{V_{in}}|$ was calculated experimentally, as well as through simulation, near the resonant peak (d=5mm).

Fig. 17 shows measured and simulated results for the case that the two coils are separated by 5mm. It is apparent from the operating frequency displayed in Fig. 17 that the measured response matches to the situational response of the circuit. Finally, power transmission efficiency through proposed distance has been analysed. The calculation of efficiency can be completed by assuming the input current and the input voltage are generating active power. Maximum measured efficiency is 38.79% at 5 mm distance. The measured output power is 772.8mW. Measured power transmission efficiency for the case that the two coils are separated by (5mm-15mm) is shown as Fig18.

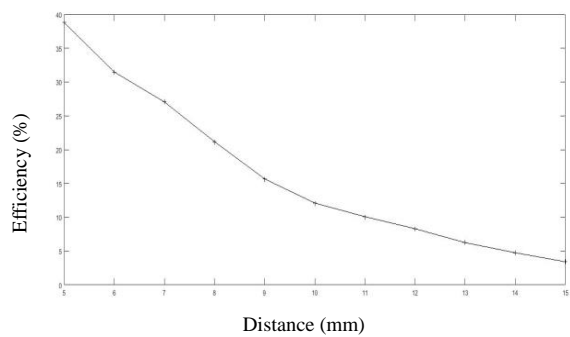

**Fig. 18.** Power transfer efficiency was measured at distances ranging from 5mm to 15mm while increasing the distance between the coils.

## 7. Conclusion

This paper has proposed a resonant inductive coupling system with printed spiral coils along with its design procedure. The comparison of experimental output voltage with the simulation ones shows an error of 20.14% at 5mm. This work has provided the fundamental concept of resonant near-field power transmission as well as the relationship between the induced voltage and the air gap along the transceiver. Using full bridge rectifier at receiver output, the small devices as the mobile phone can be easily charged.